\documentclass[12pt]{iopart}
\usepackage{mathrsfs}
\usepackage{graphicx,color}

\usepackage{iopams}
\def\ra{\rangle}

\def\dag{\dagger}

\newcommand{\ket}[1]{\left| #1 \right\rangle}

\def\bege{\begin{equation}}
\def\ende{\end{equation}}
\def\begen{\begin{eqnarray}}
\def\enden{\end{eqnarray}}

\begin{document}

\title[]{Exact ground state of the sine-square deformed XY spin chain}

\author{Hosho Katsura}

\address{Department of Physics, Gakushuin University, Mejiro, Toshima-ku, Tokyo 171-8588, Japan}
\ead{
\mailto{hosho.katsura@gakushuin.ac.jp},
}
\begin{abstract}
We study 
the sine-square deformed quantum XY chain with open boundary conditions, 
in which the interaction strength at the position $x$ in the chain of length $L$ 
is proportional to the function $f_x = \sin^2 [\frac{\pi}{L}(x-\frac{1}{2})]$. 
The model can be mapped onto a free spinless fermion model 
with site-dependent hopping amplitudes and on-site potentials 
via the Jordan-Wigner transformation. 
Although the single-particle eigenstates of this system cannot be obtained in closed form, 
it is shown that the many-body ground state is identical to that of the uniform XY chain with 
periodic boundary conditions.  
This proves a conjecture of Hikihara and Nishino 
[Hikihara T and Nishino T 2011 {\it Phys. Rev. B} \textbf{83} 060414(R)] 
based on numerical evidence. 

\end{abstract}

\submitto{\JPA}
\maketitle

\section{Introduction}
The effect of boundary conditions on the thermodynamic properties 
is usually believed to be negligible. 
It is, however, well known that the boundary conditions make a difference 
in the extrapolation procedure from finite-size results to the thermodynamic ones. 
For instance, a leading finite-size corrections to the ground state energy of a system 
with open boundary conditions (OBC) is in general larger than that  
of a system with periodic boundary conditions (PBC). 
Furthermore, OBC often leads to boundary oscillations such as Friedel oscillations. 
To suppress the finite-size effect due to the presence of open edges, the PBC 
are more often used than the OBC. 
However, in numerical studies, in particular, in the density matrix 
renormalization group (DMRG) method which is a powerful method to 
study one-dimensional (1D) strongly correlated systems, 
the OBC are more preferable than the PBC. 
This is because, although there are variants of DMRG that can work with PBC~\cite{Verstraete, Pippan},  
they require additional computational resources compared with the conventional scheme with OBC. 
There is a way to remove the boundary effects in the DMRG scheme 
by turning off the interactions smoothly around the edges~\cite{Vekic1, Vekic2}. 

Recently, a variant of the scheme for the smooth boundary condition, 
which is called the sine-square deformation (SSD), has been proposed~\cite{Gendiar}.  
In this scheme, the interaction energy scale at the position $x$ in the chain 
of length $L$ is modified according to the following function:
\begin{equation}
f_x = \sin^2 \left[ \frac{\pi}{L} \left( x-\frac{1}{2} \right) \right], 
\label{eq:SSD_func}
\end{equation} 
where $0 < x \le L$. 
In~\cite{Gendiar}, 1D free fermion systems with SSD 
were examined and it was found that the SSD completely suppresses 
the boundary effect in the ground state, i.e., the finite-size correction 
to the ground state energy is identical to that of the chain with PBC. 
Furthermore, the authors found that local correlation functions such as 
the bond strength are  translationally invariant, which indicates that 
the ground states of PBC and OBC with SSD are extremely close 
to each other. 
A class of generalizations of SSD called {\it sinusoidal} deformation 
has been applied to the free fermion systems as well as interacting 
systems such as the extended Hubbard model~\cite{Gendiar2}. 
It was again found that the leading finite-size correction to the ground-state energy 
has the same form as that of the uniform chain with PBC. 
Motivated by these observations, Hikihara and Nishino have studied 
the SSD in several 1D quantum spin systems at criticality such as the XXZ 
spin-$\frac{1}{2}$ Heisenberg chain in a field~\cite{Hikihara-Nishino}. 
They examined the ground-state properties such as entanglement entropy 
and correlation functions numerically. 
Their results strongly suggest that the ground state of an open chain 
with SSD is 
extremely close to that of a periodic chain in which the interactions 
are translation invariant. 
This conclusion was further supported by examining the wave-function overlap, 
which is very close to unity. In particular, for the XY case ($\Delta=0$), 
the wave-function overlap is exactly $1$ within the numerical accuracy. 
Note that the deviation of the overlap from the unity becomes of the order of 
$10^{-3}$ when $\Delta \ne 0$ and hence it is unlikely that the correspondence 
between the periodic chain and the open chain with SSD is exact except for $\Delta=0$. 

The aim of this communication is to prove this correspondence analytically for the XY case. 
Although the model can be mapped onto the free spinless fermion system via the Jordan-Wigner transformation, 
hopping amplitudes and on-site potentials are position dependent. 
To our knowledge, the single-particle eigenstates of this free-fermion system 
have not been obtained in closed form,  similarly to the Hofstadter problem~\cite{Hofstadter}.  
This is in contrast to the known inhomogeneous XY models whose single-particle states can be 
obtained analytically~\cite{Feldman1, Feldman2, Jeugt1}. 
Nevertheless, we shall show that the Slater determinant of plane waves in the periodic chain 
is the exact ground state of the open chain with SSD. 
To this aim, we first consider the XY chain with cosine deformation, in which 
the interaction energy scale at the position $x$ is modified by the function 
$\cos [\frac{2\pi}{L} (x-\frac{1}{2})]$. 
Then we show that the Slater determinant state is a zero-energy state of this model. 
With a simple trigonometric identity, we find that the cosine-deformed Hamiltonian 
consists of the Hamiltonian for the periodic chain and that for the open chain with SSD. 
Since the Slater determinant state is an exact eigenstate of the periodic chain, 
it is obvious that this state is also an eigenstate of the SSD Hamiltonian. 
From the Perron-Frobenius theorem, we can conclude that this state is 
the unique ground state of the SSD Hamiltonian. 

\section{Definition of the XY spin chain in a field with sine-square deformation}
The XY spin chain in a magnetic field with PBC is described by the following Hamiltonian:
\begin{equation}
{\cal H}_{\rm XY} = -J \sum^{L-1}_{j=1} (S^x_j S^x_{j+1} + S^y_j S^y_{j+1})
-J (S^x_L S^x_1 + S^y_L S^y_1)
-h\sum^L_{j=1} S^z_j, 
\label{eq:Ham_periodic}
\end{equation}
where $-J < 0$, $L$ is the total length of the chain, and $S^\alpha_j$ ($\alpha=x,y,z$) 
are spin-$\frac{1}{2}$ operators defined on the $j$th site. 
This model can be solved exactly using the Jordan-Wigner transformation~\cite{Lieb-Schultz-Mattis, Katsura}. 
It has received considerable attention for a long time and various quantities 
such as the correlation functions~\cite{McCoy68, Barouch-McCoy2, Barouch-McCoy4, Its-Izergin-Korepin-Slavnov92}, 
the emptiness formation probability~\cite{Shiroishi-Takahashi-Nishiyama01, Franchini-Abanov05}, 
and the entanglement entropy~\cite{Jin_Korepin_04, Peschel, Its_Jin_Korepin_05, Franchini_Its_Jin_Korepin_07} have been calculated exactly. 
The Hamiltonian with positive exchange $J$ can be obtained from ${\cal H}_{\rm XY}$ 
by the unitary transformation $U = \prod_{j:{\rm odd}} \exp({\bf i} \pi S^z_j)$: 
\begin{equation}
\fl~~~~~~~~~~~~
U {\cal H}_{\rm XY} U^\dag = J \sum^{L-1}_{j=1} (S^x_j S^x_{j+1} + S^y_j S^y_{j+1})
+(-1)^L J (S^x_L S^x_1 + S^y_L S^y_1)
-h\sum^L_{j=1} S^z_j,
\end{equation}
where the case of even $L$ corresponds to the PBC while that of odd $L$ corresponds 
to the antiperiodic boundary condition (APBC) (see footnote 18 of Ref.~\cite{Hikihara-Nishino}). 

Let us next introduce the Hamiltonian for the XY chain with the SSD~\cite{Hikihara-Nishino}:
\begin{equation}
{\cal H}_{\rm sXY} = -J \sum^{L-1}_{j=1} f_{j+\frac{1}{2}} (S^x_j S^x_{j+1} + S^y_j S^y_{j+1})
-h \sum^L_{j=1} f_j S^z_j, 
\label{eq:Ham_SSD}
\end{equation}
where $f_x$ is introduced in Eq. (\ref{eq:SSD_func}). 
As is obvious, this model is defined on the open chain of length $L$, i.e., 
there is no interaction between the end sites ($1$ and $L$). 
This model can also be mapped onto the free spinless fermion system via 
the Jordan-Wigner transformation. However, in contrast to the case of (\ref{eq:Ham_periodic}), 
the hopping amplitudes and on-site potentials result in position dependent. 
Therefore, the standard Fourier transform cannot be applied to solve the problem. 

Before going into details, we give a brief synopsis on the known results of 
the uniform XY chain with PBC, which will be of use later on. 
The Hamiltonian ${\cal H}_{\rm XY}$ can be written in terms of spinless fermions $c_j$ as
\begin{equation}
\fl~~~~~~~~~~~
{\cal H}_{\rm XY} = -\frac{J}{2} \sum^{L-1}_{j=1} (c^\dag_j c_{j+1} + c^\dag_{j+1} c_j)
+\frac{J}{2} \Gamma (c^\dag_L c_{1} + c^\dag_{1} c_L)
-h \sum^L_{j=1} \left( c^\dag_j c_j -\frac{1}{2} \right),
\label{eq:Ham_fermion}
\end{equation}
where the Jordan-Wigner fermions are defined through
\begin{eqnarray}
S^+_j = c^\dag_j \prod^{j-1}_{i=1} (1-2c^\dag_i c_i),~~
S^-_j = c_j \prod^{j-1}_{i=1} (1-2c^\dag_i c_i),~~
S^z_j = c^\dag_j c_j -\frac{1}{2}, 
\end{eqnarray}
with $S^\pm_j = S^x_j \pm {\bf i} S^y_j$. 
The operator $\Gamma$ in Eq. (\ref{eq:Ham_fermion}) is given by
$
\Gamma := \prod^L_{j=1} (-\sigma^z_j), 
$
which commutes with the Hamiltonian ${\cal H}_{\rm XY}$. 
Here, $\sigma^z_j$ is the Pauli matrix and related to $S^z_j$ through $\sigma^z_j =2 S^z_j$. 
Since $\Gamma^2=1$, the eigenstates of ${\cal H}_{\rm XY}$ are separated into 
two disconnected sectors with $\Gamma=\pm 1$, in which $+/-$ characterizes 
configurations with an even/odd number of up spins. 
Therefore, we have
\begin{equation}
{\cal H}_{\rm XY} = \frac{1+ \Gamma}{2} {\cal H}^+_{\rm XY} + \frac{1-\Gamma}{2} {\cal H}^-_{\rm XY},
\end{equation}
where 
\begin{equation}
\fl~~~~~~~~~~
{\cal H}^\pm_{\rm XY} = -\frac{J}{2} \sum^{L-1}_{j=1} (c^\dag_j c_{j+1} + c^\dag_{j+1} c_j)
\pm \frac{J}{2} (c^\dag_L c_{1} + c^\dag_{1} c_L)-h \sum^L_{j=1} \left( c^\dag_j c_j -\frac{1}{2} \right).
\end{equation}
Now we have to take care of the boundary conditions on the fermions. 
For $\Gamma=+1$, we have to impose APBC on the fermions 
and they have ``half-integer" quasimomenta: 
\begin{equation}
p= \frac{2\pi}{L} \left( \ell+ \frac{1}{2} \right),~~(\ell \in \mathbb{Z}),
\end{equation} 
where $-\frac{L}{2} \le \ell \le \frac{L}{2}-1$ for even $L$ and 
$-\frac{L}{2}+\frac{1}{2} \le \ell \le \frac{L}{2}-\frac{1}{2}$ for odd $L$.  
We call this sector the NS (Neveu-Schwarz) sector in which there is an even number of fermions. 
On the other hand, for $\Gamma=-1$, we require PBC 
on the fermions and they have ``integer" quasimomenta: 
\begin{equation}
p= \frac{2\pi}{L} \ell,~~(\ell \in \mathbb{Z}),
\end{equation}  
where $-\frac{L}{2} \le \ell \le \frac{L}{2}-1$ for even $L$ and 
$-\frac{L}{2}+\frac{1}{2} \le \ell \le \frac{L}{2}-\frac{1}{2}$ for odd $L$. 
We call this sector the R (Ramond) sector in which there is an odd number of fermions. 

\section{Ground state wavefunction of the uniform XY chain}
Our purpose is to show that the ground state of ${\cal H}_{\rm XY}$ is identical 
to that of ${\cal H}_{\rm sXY}$. We shall first summarize the ground state 
wavefunction of the uniform XY chain. 
Since the magnetization $m=\frac{1}{L} \sum^L_{j=1} S^z_j$ is conserved (in both ${\cal H}_{\rm XY}$ and ${\cal H}_{\rm sXY}$), 
it is convenient to work at fixed $m$. 
In finite systems, $m$ and the number of fermions (up spins) $N$ are related to each other through 
$
m=\frac{N}{L} - \frac{1}{2}. 
$
In the ground state of ${\cal H}_{\rm XY}$, one can show that, 
for a given $m$, the magnetic field should satisfy 
\begin{equation}
-J \cos \left( \frac{N}{L} \pi \right) \le h < -J \cos \left( \frac{N+1}{L} \pi \right). 
\label{eq:inequi_mag}
\end{equation}
This gives the equilibrium relation between magnetization and field in the thermodynamic limit:
\begin{equation}
h_m = J \sin (m \pi). 
\label{eq:equi_mag}
\end{equation}
In the XY chain with SSD, we restrict $h$ to the above $h_m$ 
instead of the generic one in Eq. (\ref{eq:inequi_mag}) to ensure the solvability of the model. 
The difference becomes negligible in the thermodynamic ($L \to \infty$) limit. 
Note that a similar strategy has been used in the numerical studies (see footnote 4 of Ref.~\cite{Hikihara-Nishino}). 

For fixed magnetization, the unnormalized ground-state wavefunction of ${\cal H}_{\rm XY}$ is written as
\begin{equation}
\ket{\Psi_{0}} = \sum_{\{x_1, ..., x_N \}} \det[e^{{\bf i}p_i x_j}]_{i,j=1,2,...,N} \ket{x_1, x_2, ..., x_N},
\label{eq:wf1}
\end{equation}
where $\ket{x_1, x_2, ..., x_N}$ ($1 \le x_1 < x_2 < \cdots < x_N \le L $) 
denote the configurations of up spins at $(x_1, x_2, ..., x_N)$. 
Here, the sum extends over all possible spin configurations. 
In the ground state, 
the set of quasimomenta corresponding to $m$ ($N$) is given by
\begin{equation}
p_i = \frac{2 \pi}{L} \left( \frac{N}{2} -i +\frac{1}{2} \right), 
\end{equation}
with $i=1,2, ..., N$ for both the NS and R sectors. 
Then introducing a new set of variables 
\begin{equation}
z_j := \exp\left( {\bf i} \frac{2 \pi}{L} x_j \right), 
\label{eq:zjxj}
\end{equation}
$\ket{\Psi_{\rm 0}}$ in Eq. (\ref{eq:wf1}) is rewritten as
\begin{equation}
\ket{\Psi_{0}} = \sum_{\{x_1, ..., x_N \}} 
\left( \prod^N_{i=1} z_i \right)^{-\frac{N-1}{2}} \prod_{1 \le i < j \le N} (z_i-z_j)
\ket{x_1, x_2, ..., x_N},
\label{eq:wf2}
\end{equation}
where we have used the property of the Vandermonde determinant. 
A straightforward calculation gives the ground-state energy as
\begin{equation}
E_{0} = \frac{hL}{2} - \sum^N_{i=1} (h+J \cos p_i) = h \left( \frac{L}{2} -N \right) -J \frac{\sin(\pi N/L)}{\sin (\pi/L)}.  
\end{equation}
Using the magnetization $m$, this ground state energy can be recast as
\begin{equation}
\fl~~~~~~~~
E_{0} = -h m L -\frac{J \cos (m \pi)}{\sin (\pi/L)} 
= -\left( hm+\frac{J \cos(m \pi)}{\pi} \right) L - \frac{\pi J \cos (m \pi)}{6L} + O \left(\frac{1}{L^3} \right). 
\end{equation}
The leading finite-size correction to the ground-state energy 
is in agreement with the general prediction of the $c=1$ conformal field theory~\cite{Bloete86, Affleck86}. 

\section{The XY spin chain with cosine-deformation}
To prove the conjecture of Hikihara and Nishino~\cite{Hikihara-Nishino} 
that the state $\ket{\Psi_{\rm 0}}$ defined in Eq. (\ref{eq:wf2}) is the ground state of ${\cal H}_{\rm sXY}$, it is more convenient to consider the XY model with cosine deformation. 
The Hamiltonian of this system is constructed from ${\cal H}_{\rm XY}$ and ${\cal H}_{\rm sXY}$ as
\begin{eqnarray} 
\fl~~~~~
{\cal H}_{\rm cos} &=& \frac{1}{2}{\cal H}_{\rm XY}-{\cal H}_{\rm sXY}
\label{eq:con_Ham}
\nonumber \\
\fl~~~~~
&=& -\frac{J}{4} \sum^L_{j=1} \cos \left( \frac{2 \pi}{L} j \right) (S^+_j S^-_{j+1} + S^-_j S^+_{j+1})
-\frac{h}{2} \sum^L_{j=1} \cos \left[ \frac{2 \pi}{L} \left( j-\frac{1}{2} \right) \right] S^z_j.
\end{eqnarray} 
We have another decomposition of ${\cal H}_{\rm cos}$:
\begin{equation}
{\cal H}_{\rm cos} = \frac{1}{2} ({\cal H}^+_{\rm chiral} + {\cal H}^-_{\rm chiral} ),
\end{equation}
where ${\cal H}^\pm_{\rm chiral}$ is defined by
\begin{equation}
{\cal H}^\pm_{\rm chiral} = 
-\frac{J}{4} \sum^L_{j=1} q^{\pm j} (S^+_j S^-_{j+1} + S^-_j S^+_{j+1})
-\frac{h}{2} \sum^L_{j=1} q^{\pm (j-1/2)} S^z_j, 
\end{equation}
with $q=\exp ({\bf i} \frac{2\pi}{L})$. 
We shall consider the eigenvalue problem 
${\cal H}^\pm_{\rm chiral} \ket{\Psi} = E^\pm \ket{\Psi}$
instead of that of ${\cal H}_{\rm cos}$. 
Note that ${\cal H}^\pm_{\rm chiral}$ are non-Hermitian and their 
eigenvalues may not be real. 
For fixed magnetization $m$ and hence fixed number of up spins ($N$), 
the eigenstate can be written in the form 
\begin{equation}
|\Psi \ra = \sum_{\{x_1, ..., x_N \}} \Psi (z_1, z_2, ..., z_N) \ket{x_1, x_2, ..., x_N}, 
\end{equation}
where the coordinates $\{ x_1, x_2, ..., x_N \}$ should be distinct due to the hard-core constraint. 
Let us now derive the Hamiltonians in the first quantized form. 
To do so, we introduce  the $q$-shift operator $T_{q,j}$ 
which acts on a function of $\{ z_j \}^N_{j=1}$ as
\begin{equation}
T_{q,j} F(z_1, ..., z_{j-1}, z_j, z_{j+1},...,z_N) = F(z_1, ..., z_{j-1}, q z_j, z_{j+1},...,z_N). 
\end{equation} 
Note that $F (\{ z_j \}^N_{j=1})$ is restricted to be a function such that 
$F (\{ z_j \}^N_{j=1})=0$ when any two out of $N$ coordinates coincide. 
Using $T_{q, j}$, the action of ${\cal H}^\pm_{\rm chiral}$ on $\Psi (z_1, z_2, ..., z_N)$ 
is written as $H^\pm_{\rm chiral} \Psi (z_1, z_2, ..., z_N) = E^\pm \Psi (z_1, z_2, ..., z_N)$ with
\begin{eqnarray}
H^+_{\rm chiral} = -\frac{J}{4} \sum^N_{j=1} z_j (T_{q,j}+q^{-1} T_{q^{-1},j})
-\frac{h_m}{2} \sum^N_{j=1} q^{-1/2} z_j, \\
H^-_{\rm chiral} = -\frac{J}{4} \sum^N_{j=1} z^{-1}_j (T_{q, j} + q T_{q^{-1},j}) 
-\frac{h_m}{2} \sum^N_{j=1} q^{1/2} z^{-1}_j,
\end{eqnarray}
where we have used the fact that $\sum^L_{j=1} q^j = \sum^L_{j=1} q^{-j} = 0$. 
Note that the magnetic field has been specified according to 
the relation defined in Eq. (\ref{eq:equi_mag}). 
The model is no longer solvable even at the field with a small deviation from $h_m$. 
This is because the Zeeman term with the cosine deformation (or its chiral counterpart) 
does not commute with the exchange term with the same deformation. 

We shall show that the Slater determinant state
\begin{equation}
\Psi_{0} ( z_1, z_2, ..., z_N) = \left( \prod^N_{i=1} z_i \right)^{-\frac{N-1}{2}} \prod_{1 \le i < j \le N} (z_i-z_j)
\label{eq:gswf}
\end{equation}
is the exact zero-energy state of $H^\pm_{\rm chiral}$. 
Using a similarity transformation, we have
\begin{eqnarray}
\fl~~~~~ 
&& [\Psi_{0} (\{ z_j \}^N_{j=1})]^{-1} H^+_{\rm chiral} \Psi_{0} (\{ z_j \}^N_{j=1})  
\nonumber \\
\fl~~~~~ 
&=& -\frac{J}{4}\sum^N_{j=1} \Bigg( 
 q^{-\frac{N-1}{2}} z_j \prod_{k (\ne j)} \frac{q z_j-z_k}{z_j-z_k}
+ q^{\frac{N-1}{2}} q^{-1}z_j \prod_{k (\ne j)} \frac{q^{-1} z_j-z_k}{z_j-z_k} \Bigg)
\nonumber \\
\fl~~~~~
&& + \frac{J}{4} \sum^N_{j=1} (q^{N/2}+q^{-N/2}) q^{-1/2} z_j, 
\label{eq:similarity}
\end{eqnarray}
where we have used the relation (\ref{eq:equi_mag}). 
Note that $\Psi_{0} (\{ z_j \}^N_{j=1})$ is nonvanishing for any configuration 
$\{ z_j \}^N_{j=1}$, which will be shown in the next section. 
Using the following identity 
\begin{equation}
\sum^N_{j=1} z_j \prod_{k (\ne j)} \frac{z_j-t z_k}{z_j-z_k} = \sum^N_{j=1} z_j,
\label{eq:identity}
\end{equation}
where $t$ is an arbitrary complex number, 
we can further simplify Eq. (\ref{eq:similarity}). 
As a result, we obtain $[\Psi_{0} (\{ z_j \})]^{-1} H^+_{\rm chiral} \Psi_{0} (\{ z_j \}) =0$, 
which means $H^+_{\rm chiral} \Psi_{0} (\{ z_j \}) =0$. 
Along the same lines as above, we can also show that $H^-_{\rm chiral} \Psi_{0} (\{ z_j \}) =0$ using the identity Eq. (\ref{eq:identity}).  Therefore, we arrive at
\begin{equation}
H_{\rm cos} \Psi_0 (z_1, z_2, ..., z_N) = \frac{1}{2} (H^+_{\rm chiral} + H^-_{\rm chiral}) \Psi_0 (z_1, z_2, ..., z_N) =0,
\end{equation}
where $H_{\rm cos}$ is the first quantized Hamiltonian for ${\cal H}_{\rm cos}$. 
Therefore, $\ket{\Psi_{0}}$ in Eq. (\ref{eq:wf2}) is the exact zero-energy state of ${\cal H}_{\rm cos}$. 
It should be noted that $\ket{\Psi_{0}}$ is not necessarily the ground state of ${\cal H}_{\rm cos}$. 
Since $\ket{\Psi_{\rm 0}}$ is the eigenstate of ${\cal H}_{\rm XY}$ with 
the eigenenergy $E_{\rm 0}$, we immediately arrive at
\begin{equation}
{\cal H}_{\rm sXY} \ket{\Psi_{0}} = \frac{E_{0}}{2} \ket{\Psi_{0}},
\end{equation}
where we have used the relation among ${\cal H}_{\rm cos}$, ${\cal H}_{\rm XY}$, and ${\cal H}_{\rm sXY}$ (Eq. (\ref{eq:con_Ham})). 
Therefore, $\ket{\Psi_{0}}$ is an exact eigenstate of the XY model in the field with SSD. 

Several comments are in order. 
The remarkable identity Eq. (\ref{eq:identity}) can be directly proved 
by comparing the residues of both sides. 
Another way of obtaining Eq. (\ref{eq:identity}) is to make use of the generating function, 
which is used to construct Macdonald's $q$-difference operators whose eigenfunctions 
are the Macdonald polynomials (see Sec VI in \cite{Macdonald}). 
The generating function is defined through
\begin{equation}
\prod^n_{j=1} \frac{1-t z_j w}{1- z_j w} = \sum_{r \ge 0} g_r (z_1, ..., z_n;0, t) w^r,
\label{eq:generating_function}
\end{equation}
where $g_0 (z_1, ..., z_n; 0, t)=1$ and 
\begin{equation}
g_r (z_1, ..., z_n; 0, t) = (1-t) \sum^n_{j=1} z^r_j \prod_{k (\ne j)} \frac{z_j -t z_k}{z_j - z_k}~~~~~(r \ge 1). 
\end{equation}
By comparing the coefficients of $w^1$ of both sides in Eq. (\ref{eq:generating_function}), 
one can obtain Eq. (\ref{eq:identity}). 
An interesting corollary from the fact ${\cal H}^\pm_{\rm chiral} \ket{\Psi_0}=0$ is that
$\ket{\Psi_0}$ is also a zero-energy state of the XY Hamiltonian with sine-deformation 
defined by
\begin{eqnarray}
\fl~~~~~
{\cal H}_{\rm sin} &=& \frac{1}{2{\bf i}} ({\cal H}^+_{\rm chiral} - {\cal H}^-_{\rm chiral}) \nonumber \\
\fl~~~~~
&=& -\frac{J}{4} \sum^L_{j=1} \sin \left( \frac{2 \pi}{L} j \right) (S^+_j S^-_{j+1} + S^-_j S^+_{j+1})
-\frac{h}{2} \sum^L_{j=1} \sin \left[ \frac{2 \pi}{L} \left( j-\frac{1}{2} \right) \right] S^z_j.
\end{eqnarray}
When $h=0$, the free-fermion system corresponding to ${\cal H}_{\rm sin}$ is 
related to the Hofstadter (or Harper) problem.  
This indicates that one can explicitly write down the many-body eigenstate of 
the Hofstadter problem on a thin torus~\cite{Katsura-Maruyama-Tanaka-Tasaki} at half-filling 
even though one cannot obtain the single-particle eigenstates in closed form 
except when the single-particle energy is zero~\cite{Hatsugai-Kohmoto-Wu}. 
Many other variants can be constructed from ${\cal H}^+_{\rm chiral}$ 
and ${\cal H}^-_{\rm chiral}$, in which $\ket{\Psi_0}$ is the exact zero-energy state. 
It is worth mentioning that the explicit definition of $q$ is not necessary for the proof 
of the zero-energy state. 
Instead, we have used the relation between $z_j$ and $x_j$, i.e., $z_j = q^{x_j}$ (see Eq. (\ref{eq:zjxj})). 
We can, therefore,  apply the above argument to a large family of models 
including the systems with the hyperbolic deformation~\cite{Ueda-Nishino08, Ueda-Gendiar10,  Ueda-Nakano10} 
and those with the exponential deformation~\cite{Okunishi-Nishino10}, in which 
the parameter $q$ is taken to be real. 

\section{Uniqueness of the ground state}
So far, we have shown that Slater determinant state $\ket{\Psi_{\rm 0}}$ is the exact eigenstate of ${\cal H}_{\rm sXY}$. 
In this section, we shall prove that $\ket{\Psi_{0}}$ is the unique ground state of ${\cal H}_{\rm sXY}$. 
For fixed magnetization, we specify the basis state we work with as 
$\ket{x_1, x_2, ..., x_N}$, where $1 \le x_1 < x_2 <  ... < x_N \le L$ denote the positions of up spins. 
In this basis, all of the off-diagonal matrix elements of ${\cal H}_{\rm sXY}$ 
(Eq. (\ref{eq:Ham_SSD})) are non-positive. 
Since $f_{j+\frac{1}{2}} >0$ for all $j=1, 2, ..., L-1$, it is obvious that ${\cal H}_{\rm sXY}$ 
satisfies the connectivity condition. 
Therefore, the Perron-Frobenius theorem applies and hence the ground state 
is nondegenerate and can be written as a linear combination of 
all the basis states: 
\begin{equation}
\ket{\Psi_{\rm gs}} = \sum_{\{x_1, ..., x_N \}} 
c_{x_1, x_2, ..., x_N}
\ket{x_1, x_2, ..., x_N},
\label{eq:canonical_form}
\end{equation}
where the coefficients $c_{x_1, x_2, ..., x_N}$ are strictly positive for any $(x_1, x_2, ..., x_N)$. 
Let us examine whether the state $\ket{\Psi_0}$ in Eq. (\ref{eq:wf2})
has the same property. 
Using the relation (\ref{eq:zjxj}), 
one can write down the component of $\ket{\Psi_0}$ as
\begin{equation}
\Psi_0 (z_1, ..., z_N) = (-2 {\bf i})^{\frac{N (N-1)}{2}} \prod_{1 \le i < j \le N} 
\sin \left( \frac{\pi}{L} (x_j - x_i) \right). 
\end{equation}
Therefore, the state 
\begin{equation}
|{\tilde \Psi}_0 \rangle =  (-2 {\bf i})^{-\frac{N (N-1)}{2}} \ket{\Psi_0}
\end{equation}
can be expressed in the form of Eq. (\ref{eq:canonical_form}). 
Here, we have used the fact $1 \le x_j - x_i < L$ when $i < j$.  
Since there can be no other state with positive coefficients only that is orthogonal 
to the ground state, $\ket{\Psi_{0}}$ is identical to $\ket{\Psi_{\rm gs}}$ apart from an  
overall factor. This proves that $\ket{\Psi_0}$ is the unique ground state of ${\cal H}_{\rm sXY}$.  

\section{Concluding remarks}
\label{sec:}
In this communication, we have studied the XY chain in a field with SSD. 
We have shown that the unique ground state of this model is identical to 
that of the uniform XY chain with PBC in each sector of fixed magnetization. 
This explains the previous observations in numerical 
studies that the SSD completely suppresses the effect of open boundaries 
on the ground state of the free-fermion or XY chain~\cite{Gendiar, Gendiar2, Hikihara-Nishino}. 
We also found that the simultaneous ground state of the uniform and SSD XY chains 
is the zero-energy state of a large class of Hamiltonians including the XY chain 
with sine/cosine deformation. 
The identity (\ref{eq:identity}) playing a crucial role in the proof 
has its origin in Macdonald's $q$-difference operators. 
We can generate an infinite number of identities from Eq. (\ref{eq:generating_function}). 
Therefore, it is interesting to ask whether these identities give 
a new class of models whose ground states can be obtained exactly. 
In contrast to the XY (free-fermion) case, 
previous works~\cite{Hikihara-Nishino} suggest that 
it is unlikely that the correspondence between the periodic chain 
and the open chain with SSD is exact in the general XXZ chain. 
However, this does not exclude the possibility of another deformed Hamiltonian  
whose ground state is exactly identical to that of the periodic XXZ chain.  
In this respect, it is worth recalling that the ground state of the XXZ chain 
in the antiferromagnetic regime is also an eigenstate of the corner transfer 
matrix for the six-vertex model~\cite{Frahm-Thacker91}. 
Although the corner transfer matrix is well-defined only in the massive regime, 
the models with hyperbolic deformations have some similarities with the 
corner Hamiltonian~\cite{Ueda-Nishino08}. 

\ack{}
The author would like to thank 
Toshiya Hikihara, Isao Maruyama, and Tomotoshi Nishino  
for their valuable comments and suggestions. 

\section*{References}
\providecommand{\newblock}{}

\end{document}